# Switchable photovoltaic and polarization modulated rectification in Si-integrated Pt/(Bi$_{0.9}$Sm$_{0.1}$)(Fe$_{0.97}$Hf$_{0.03}$)O$_3$/LaNiO$_3$ heterostructures

Radhe Agarwal, Yogesh Sharma, and Ram S. Katiyar

*Department of Physics and Institute for Functional Nanomaterials, University of Puerto Rico, San Juan, PR 00931, USA*

**Abstract**

We studied switchable photovoltaic and photo-diode characteristics of Pt/(Bi$_{0.9}$Sm$_{0.1}$)(Fe$_{0.97}$Hf$_{0.03}$)O$_3$/LaNiO$_3$ (Pt/BSFHO/LNO) heterostructures integrated on Si (100). The directions of photocurrent (J$_{SC}$) and rectification are found to be reversibly switchable after applying external poling voltages. In pristine state, metal-ferroelectric-metal capacitor Pt/BSFHO/LNO shows J$_{SC}$ ~32 µA/cm$^2$ and V$_{OC}$ ~0.04 V, which increase to maximum value of J$_{SC}$ ~ 303 (-206) µA/cm$^2$ and V$_{OC}$ ~ -0.32 (0.26) V after upward (downward) poling at ±8 V. We believe that Schottky barrier modulation by polarization flipping at Pt/BSFHO interface could be a main driving force behind switchable photovoltaic and rectifying diode characteristics of Pt/BSFHO/LNO heterostructures.

___________________________________________

Corresponding authors email ID: rkatiyar@hpcf.upr.edu (Ram S. Katiyar), radhe.agarwal@upr.edu (Radhe Agarwal); Tel : 7877514210. Fax: 7877642571

Multiferroics are very promising materials due to their distinct physical properties i.e., occurrence of magnetic and ferroelectric ordering simultaneously in single phase and possibility to revise one by altering other due to the likely coupling between both ferroic orderings.[1-8] BiFeO$_3$ (BFO) is one of the most studied lead free room temperature multiferroic perovskite oxide because of its fascinating physical properties; large polarization (90μC/cm$^2$), weak ferromagnetism, and optical bandgap (~2.67 eV)[9-12] lies in visible region, make it suitable multifunctional material for spintronics, data storage, sensors, and optoelectronic applications.[13-17] Recently, the single phase BFO thin films showed promising potential towards ferroelectric photovoltaic (Fe-PV) application due to enhanced room temperature ferroelectricity and small optical bandgap.[17-20] Among other wide bandgap Fe-PV materials (i.e., BaTiO$_3$, LiNbO$_3$ and Pb(Zr,Ti)O$_3$), BFO exhibits large open circuit voltage ($V_{OC}$), tunable output, and switchable photodiode effect [6, 9-11]. It has been reported that the polarization flipping induced modulation of Schottky-like barrier at metal/ferroelectric interface leads switchable PV and diode effect in BFO thin film heterostructures. In leaky ferroelectric BFO thin films, the migration of positively charged oxygen vacancies under the influence of external applied field dominates over the polarization flipping effect to determine switchable PV and diode behavior. This effect of oxygen vacancies deteriorates the photovoltaic performances in terms of unstable photocurrent transient where photocurrent is found to decrease slowly over multiple cycles.[21] Moreover, the sign of photocurrent could be independent of the polarization direction when the modulation of photocurrent induced by oxygen vacancies is large enough to offset that induced by polarization.[22] In BFO films having high concentration of charge defect, it could be quite evasive to conclude the role of polarization flipping and/or oxygen vacancies on switchable PV response. To study the sole impact of polarization flipping on the PV properties, one has to be taking care of the secondary phases and charge defects in BFO films. Therefore, we are co-substituting BFO with isovalent Sm ions (at Bi-site) to reduce the stereochemical activity of 6s$^2$ lone pair imbalance of Bi-ion, and aliovalent Hf ions (at Fe-site) to suppress the charged defects and improve the ferroelectric behavior of single phase doped-BFO thin films.[23-25]

In this paper, we report the switchable photo-diode and Fe-PV effect in co-substituted $(Bi_{0.9}Sm_{0.1})(Fe_{0.97}Hf_{0.03})O_3$ (BSFHO) ferroelectric thin films obtained by constituting BFO. Highly (100)-oriented BSFHO thin films with the thickness of ~320 nm were deposited on LaNiO$_3$ (LNO)-buffered Si (100) substrate using pulsed laser deposition (PLD) technique. The deposition chamber was evacuated to a base pressure of ~10$^{-6}$ torr prior to the deposition of ~100 nm thick LNO films on the Si (100) substrate at the fixed temperature of 700 °C and in oxygen ambient at a partial pressure of ~110 mTorr. Si-substrates were immersed in 2% HF: H$_2$O solution for 30 seconds to remove the native oxide before loading to PLD chamber for LNO deposition. In the second step of deposition, the BSFHO films of thickness ~320 nm were deposited on LNO covered Si (100) substrates at the fixed temperature of 690 °C and at an oxygen partial pressure of ~80 mTorr. After deposition, films were cooled to room temperature at 10 °C/min in oxygen partial pressure of 150 mTorr. With the same set of deposition parameters, BSFHO thin films (~ 300 nm) were also deposited on optical-grade (two side polished) quartz (001) crystals for optical measurements. To fabricate metal-insulator-metal (MIM) capacitor structure, square Pt-top electrodes of thickness ~70 nm were deposited at room temperature by dc-magnetron sputtering technique through a metal shadow mask with a square area of 80 x 80 μm$^2$. Crystalline structure was studied using X-ray diffraction (XRD) technique (Rigaku D/Max Ultima III with a CuKα source of wavelength λ=1.5405 A$^o$) operated at a scan rate of 0.5$^0$/min over the angular range (2θ) of 20-80°. Ferroelectric domains switching in BSFHO films were analyzed using piezoresponse force microscopy (PFM). Macrosocpic ferroelectric properties were characterized using RT6000 loop tester (Radiant Technologies). Photovoltaic measurements were performed using Keithley-2401 electrometer under 1 sun AM 1.5 solar simulator with light source density ~ 1kW/m$^2$.

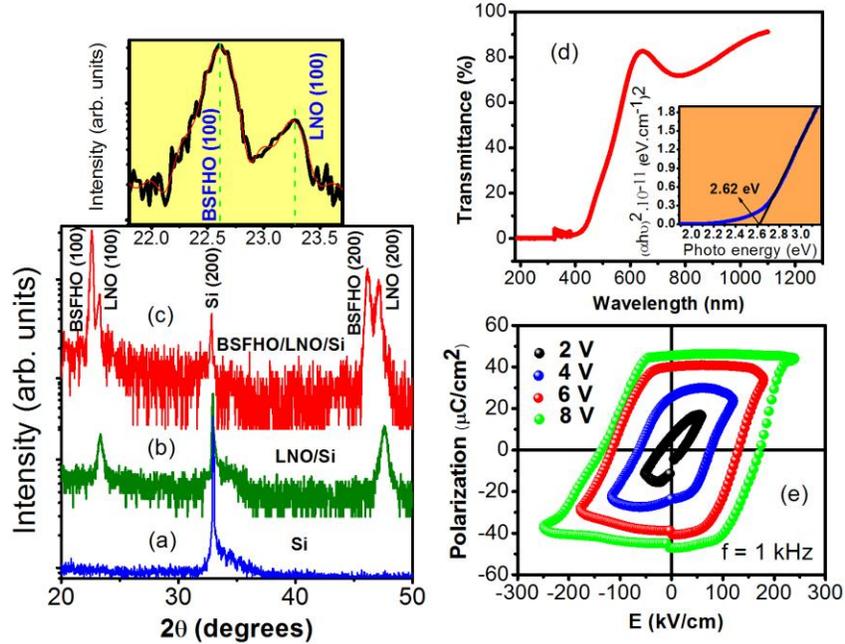

**Figure 1.** (a)-(c) XRD patterns of Si (100) substrate, LNO/Si, and BSFHO/LNO/Si heterostructures, respectively. Inset shows the zooming view of (100) peaks of BSFHO and LNO. (d) Optical transmittance of BSFHO film deposited on quartz substrate. Inset is the $(\alpha h\nu)^2$ vs $h\nu$ plot with a linear extrapolation to zero. (e) Room temperature P-E loops of Pt/BSHFO/LNO capacitor at different applied voltages and at $f$ = 1 kHz.

Figures 1(a)-(c) show the room temperature X-ray diffraction patterns of Si (100) substrate, LNO/Si, and BSFHO/LNO/Si heterostructures, respectively. Highly (100)-oriented growth of BSFHO film that appears to be single phase, is confirmed from XRD pattern, as shown in Fig. 1 (c). The (100) diffraction peak of BSFHO film is found to be slightly shifted towards higher angle than expected ($2\theta = 22.45$) for pure BFO crystal [inset of Fig. 1(c)], implying that our BSFHO film is weakly strained. For the optical band gap measurement of BSFHO thin film, UV-Visible transmittance spectrum was recorded between 1100 to 190 nm wavelength range, as shown in Fig. 1(d). The direct band gap calculation of the BSFHO films is done by a linear extrapolation of $(\alpha h\nu)^2$ versus $h\nu$ plot to zero [inset of Fig. 1(d)]. The band gap ($E_g$) of the BSFHO film was found to be ~2.62 eV, lower than the reported $E_g$ of ~2.67 eV for pure BFO films, [9-11] suggesting that BSFHO films can absorb more incident photon energy in the visible region. Figure 1(e) shows the ferroelectric polarization versus electric field (P-E) hysteresis loops of the metal-ferroelectric-metal (M–F–M) capacitor Pt/BSHFO/LNO, measured at 1 kHz

and room temperature under different applied electric voltages. We observed that at low applied electric fields, depolarization factor significantly reduce the effective field. Above an applied electric field of ~100 kV/cm, the poling behavior becomes perceptible as manifested by increase in remnant polarization with applied field. We observed well saturated P-E hysteresis loop with a maximum remnant polarization of ~46 $\mu C/cm^2$ at an applied electric field of ~160 kV/cm. Rectangular P-E hysteresis loop indicates significantly low leakage current through the BSFHO films. It has been reported that the LNO buffer layer can also help to improve the ferroelectricity in BFO films by effectively eliminate the interfacial defects.[26, 27]

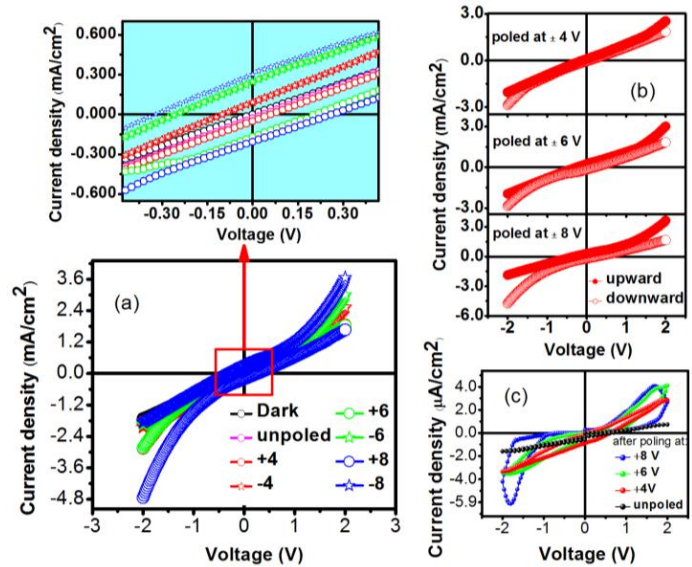

**Figure 2.** (a) J-V characteristics of virgin and poled Pt/BSFHO/LNO films under light illumination. Inset shows an enlarged view confirming the effect of poling voltages on photovoltaic characteristics. (b) Rectifying J-V characteristics at different poling voltages indicating forward/reverse photo-diode characteristics under light illumination. (c) Closed cycle J-V characteristics measured after poling at different positive voltages in dark, confirming ferroelectric resistive switching in Pt/BSFHO/LNO capacitor.

To measure the Fe-PV effect of M–F–M capacitor Pt/BSHFO/LNO, the current density–voltage (J-V) characteristics were obtained under dark and white light illumination by applying electric poling of different bias voltages of ±4, ±6, and ±8 V, as can be seen from Fig. 2(a). Both direction as well as magnitude of poling voltages have been observed to give significant contribution to the photovoltaic nature of BSFHO films. Herein, applying a negative (positive) voltage on the top Pt-electrode is defined as upward (downward) poling. Under light illumination, the J–V curves show increase in photo-current density both in upward polarization states (UPS) and downward polarization states (DPS) along with appreciable $J_{SC}$ and $V_{OC}$ values [inset of Fig. 2(a)]. We observed switchable PV effect in BSFHO films where both short circuit current density ($J_{SC}$) and open circuit voltage ($V_{OC}$) were found to switch their direction accompanying polarization flipping/switching. $J_{SC}$ direction was found to be always opposite to the polarization direction.

**Table 1**. Photovoltaic performance parameters from J-V data measured on Pt/BSFHO/LNO capacitor at different poling voltages.

| Poling Voltage | $V_{OC}$ (V) | $J_{SC}$ (µA/cm$^2$) |
|---|---|---|
| Unpoled | (+)0.05 | (-)35 |
| (+)4 V | (+)0.06 | (-)60 |
| (-)4 V | (-)0.11 | (+)102 |
| (+)6 V | (+)0.21 | (-)180 |
| (-)6 V | (-)0.25 | (+)212 |
| (+)8 V | (+)0.27 | (-)222 |
| (-)8V | (-)0.32 | (+)308 |

Table 1 shows the values of $J_{SC}$ and $V_{OC}$ at different poling voltages: at ±8 poling voltage, $J_{SC}$ ~ -205 (310) µA/cm$^2$ and $V_{OC}$ ~ 0.26 (-0.32) V were observed with downward (upward) polarization, which is almost ten times of $J_{SC}$ of unpoled sample. The improved photovoltaic properties due to electric poling indicate the intrinsic photovoltaic effect in BSFHO films where switchable photoresponse can be explained by the polarization flipping. From J-V characteristics, it is important to notice that a diode-like rectification effect accompanying switching of the rectification direction can be observed while poling the BSFHO film at the applied voltages of ≥ ±4 V. Re-plotting the J-V data in Fig. 2(b), the curves by the solid circles (upward poling) show an obvious diode-like rectification in forward bias direction, and the curves by the open circles (downward poling) indicate a reverse bias direction, under light illumination. However, the I-V curves in Fig. 2 (b) are slightly deviating from ideal diode behavior where reverse current is no longer saturated. This behavior can be modeled as an ideal diode in parallel/series with a fixed resistor component. In addition, we observed ferroelectric resistive switching in poled BSFHO films in absence of light illumination. As shown in Fig. 2(c), the closed cycle J-V characteristics of Pt/BSFHO/LNO capacitor were carried out after applying electrical poling of positive voltages. No resistance switching was observed in the J-V curve in case of unpoled BSFHO films. Whereas, a clear resistance switching can be observed in terms of J-V hysteresis after applying poling voltages of ≥ +4 V, revealing relation between current hysteresis and ferroelectric polarization. Switchable PV behaviour are further confirmed by temporal dependence of $J_{SC}$ and $V_{OC}$ with several cycles (duration 20 seconds) of light on and off under the same poling conditions, as shown in Fig. 3 (a)-(h). Both $J_{SC}$ and $V_{OC}$ showed switchable photoresponse depending on the polarization direction, with good retention and stability over time.

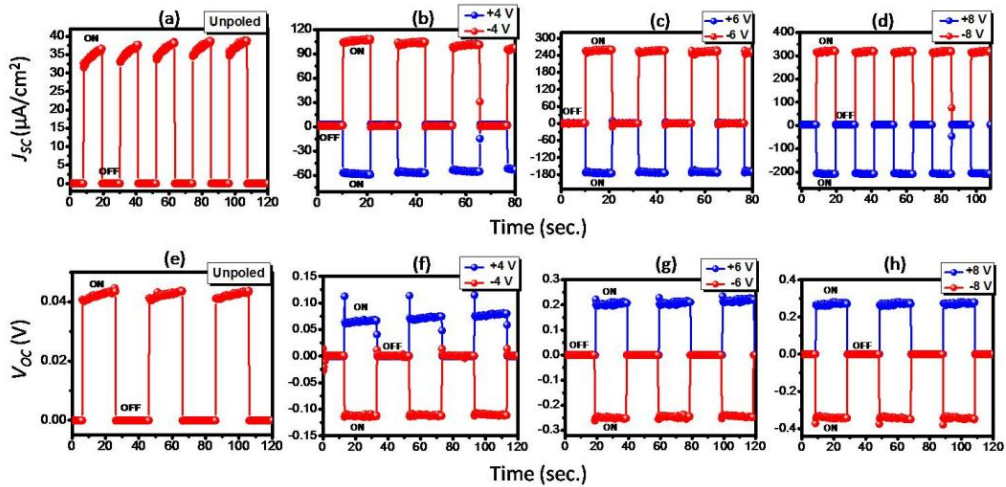

**Figure 3.** Time dependent photocurrent ($J_{SC}$) and photovoltage ($V_{OC}$) of virgin and poled Pt/BSFHO/LNO capacitors at different poling voltages with light on and off cycles of 20 seconds duration.

To justify the role of polarization flipping on switchable photoresponse in BSFHO films, we carried out PFM measurements under different applied electric bias voltages. As can be seen from Fig. 4 (a)-(c), the polarization switching can be observed from the piezoelectric phase imaging after applying different bias voltages. It can be noticed that a clear polarization switching starts to appear at bias voltages of ≥ ±4 V and becomes more obvious at ±6 and ±8 V. The polarization switching is further confirmed by the phase hysteresis loops in Fig. 4(b) and (c) for the bias voltage of ±8 V, where the measured loop show ferroelectric characteristics in terms of butterfly amplitude loops and nearly 180° phase change when the amplitude is at a minimum. Asymmetric behavior of PFM loops can be attributed to the different work functions of top/bottom electrodes.[28, 29] We believe that there can be a dominant contribution of ferroelectric polarization compared to oxygen vacancies towards switchable photoreponse and rectification in our BSFHO films due to the following facts: i) the rectification as well as photo-current directions were found to be highly dependent on the ferroelectric polarization direction, and ii) no photocurrent degradation was observed when the time dependence of $J_{SC}$ and $V_{OC}$ were measured after applying different poling voltages.[30] Therefore, polarization flipping could be the main driving force for the observed switchable PV and diode effect in Pt/BSFHO/LNO heterostructures.

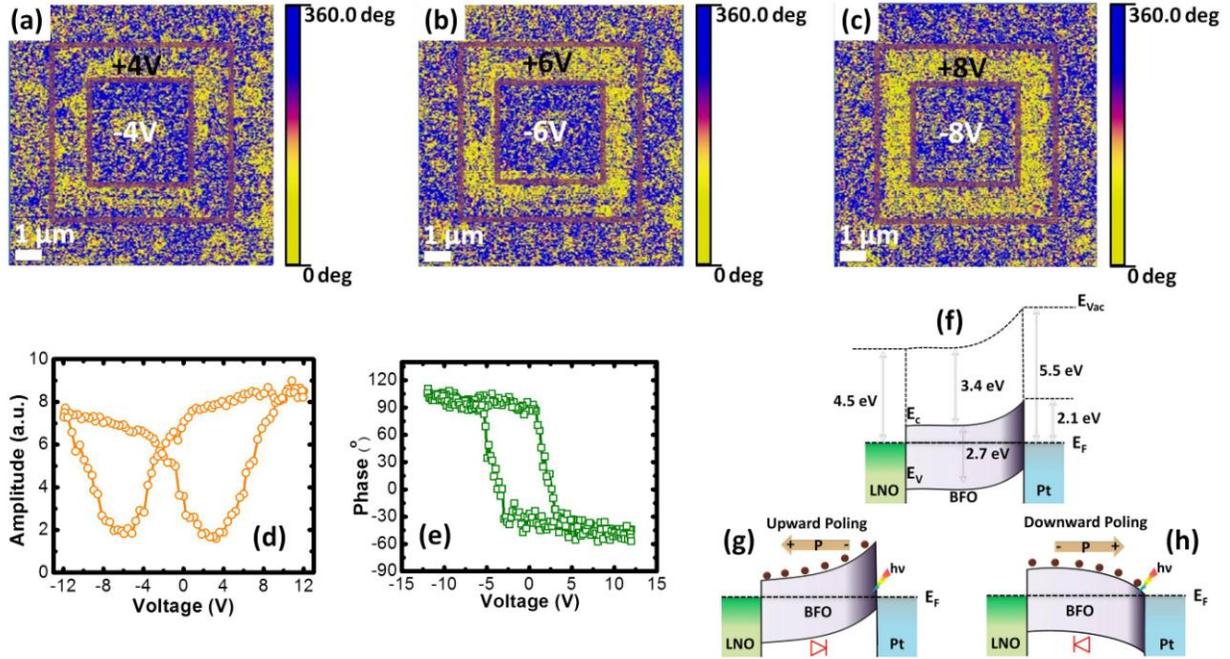

**Figure 4.** Out of plane PFM images of Pt/BSFHO/LNO capacitors at different applied electric voltages: (a) ±4 V, (b) ±6 V, and (c) ±8V. PFM (a) amplitude and (e) phase hysteresis loops. (f) Schematic representation of energy band diagram of Pt/BSFHO/LNO heterostructure at equilibrium. Band-bending at Pt/BSFHO interface in (g) upward polarization state and (h) downward polarization state.

In order to understand the effect of polarization flipping on the switchable photovoltaic response in BSFHO films, schematic energy band diagrams across the Pt/BSFHO/LNO heterostructure under unpoled and poled (up/down) conditions are represented in Fig. 4 (f)-(h). The work function (φ) of LNO and Pt are taken as, φ ~4.5 eV and 5.3 eV, respectively.[31, 32] We are considering BSFHO as weakly n-doped (φ ~4.6±1 eV and electron affinity ~3.3 eV), because complete elimination of oxygen vacancies are not possible in case of perovskite oxide thin films grown by PLD. Figure 4(f) shows the band diagram for an ideal metal–semiconductor–metal Pt/BSFHO/LNO heterostructure without electric poling. A nearly flat-band can be expected at the LNO/BSFHO interface due to the close match between work functions and possible screening of ferroelectric polarization by ionic displacement at the interface, which certainly quench the band bending effect.[18] Whereas a Schottky barrier is expected at the metal/ferroelectric (Pt/BSFHO) interface. It has been reported that a significant change in the band structure near the Pt/BFO interface can be induced by the large polarization charge.[18] In case of upward polarization, upward band bending can be formed at Pt/BSFHO interface generating a negative $V_{OC}$ and a positive $J_{SC}$, as shown in Fig. 4(g). Similarly, poling field along the polarization direction, leads unidirectional current as a forward diode. The forward bias of rectification is reversed in case of downward polarization [Fig. 4(h)], leading to a positive $V_{OC}$ and a negative $J_{SC}$, and reverse diode behavior.

In summary, highly oriented BFSHO/LNO heterostructures were deposited on Si (100) using pulsed laser deposition. Under light illumination, ferroelectric capacitor Pt/BFSHO/LNO demonstrated improved switchable photovoltaic behavior including polarization modulated rectification. Polarity dependent switchable photovoltaics, rectifying diode characteristics, ferroelectric resistive switching, and stable photoresponse over time indicated that polarization flipping could be a dominant factor over oxygen vacancies contribution to explain the mechanism behind the switchable photoresponse and diode rectification in Pt/BFSHO/LNO ferroelectric capacitor heterostructures.

**Acknowledgement:** This work was supported by the DOE EPSCoR Grant DE-FG02-08ER46526. R. A. and Y. S. acknowledge receiving graduate fellowships from NSF-IFN Grant # 1002410.

**References:**

1. J. Wang, J. B. Neaton, H. Zheng, V. Nagarajan, S. B. Ogale, B. Liu, D. Viehland, V. Vaityanatha, D. G. Schlom, U. V. Waghmare, N. A. Spaldin, K. M. Rabe, M. Wuttit, and R. Ramesh, *Science* **299**, 1719 (2003).
2. C. Michel, J.-M. Moreau, G. D. Achenbach, R. Gerson, and W. J. James, *Solid State Commun.* **7**, 701 (1969).
3. W. Eerenstein and N. D. Mathur, J. F. Scott, *Nature* **442**, 759 (2006).
4. C. H. Yang, J. Seidel, S. Y. Kim, P. B. Rossen, P. Yu, M. Gajek, Y. H. Chu, L. W. Martin, M. B. Holcomb, Q. He, P. Maksymovych, N. Balke, S. V. Kalinin, A. P. Baddorf, S. R. Basu, M. L. Scullin and R. Ramesh, *Nature Mater.* **8,** 485 (2009).
5. G. Catalan and J. F. Scott, *Adv. Mater.* **21**, 2463 (2009).
6. T. Choi, S. Lee, Y. Choi, V. Kiryukhin, and S. Cheong, Science **324**, 63 (2009).
7. A. Q. Jiang, C. Wang, K. J. Jin, X. B. Liu, J. F. Scott, C. S. Hwang, T. A. Tang, H. B. Lu, G. Z. Yang, *Adv. Mater.* **23**, 1277 (2011).
8. S. Y. Yang, L. W. Martin, S. J. Byrnes, T. E. Conry, S. R. Basu, D. Paran, L. Reichertz, J. Ihlefeld, C. Adamo, A. Melville, Y.-H. Chu, C.-H. Yang, J. L. Musfeldt, D. G. Schlom, J. W. Ager III and R. Ramesh, *Appl. Phys. Lett.* **95**, 062909 (2009).
9. W. Ji, K. Yao, and Y. Liang, *Adv. Mater.* **22**, 1763 (2010).
10. H. T. Yi, T. Choi, S. G. Choi, Y. S. Oh, and S.-W. Cheong, *Adv. Mater.* **23**, 3403 (2011).
11. S. Y. Yang, J. Seidel, S. J. Byrnes, P. Shafer1, C.-H. Yang, M. D. Rossell, P. Yu, Y.-H. Chu, J. F. Scott, J. W. Ager III, L. W. Martin, and R. Ramesh, *Nat. Nanotechnol*. **5**, 143 (2010).
12. J. Seidel, D. Fu, S. Y. Yang, O. L. Alarc, J. Wu, R. Ramesh, and J. W. Ager III, *Phys. Rev. Lett.* **107**, 126805 (2011).
13. J. A. Klug, M. V. Holt, R. N. Premnath, A. Joshi-Imre, S. Hong, R. S. Katiyar, M. J. Bedzyk, and O. Auciello, Appl. Phys. Lett. 99, 052902 (2011).
14. M. Park, S. Hong, J. A. Klug, M. J. Bedzyk, O. Auciello, K. No, and A. Petford-Long, Appl. Phys. Lett. 97, 112907 (2010).
15. M. Park, K. No, and S. Hong, AIP Advances 3, 042114 (2013).
16. "Emerging Non-volatile Memories," edited by S. Hong, O. Auciello, D. Wouters. Springer, New York, Chap. 3 (2014).

17. A. Anshul, H. Borkar, P. Singh, P. Pal, S. S. Kushvaha, and A. Kumar, Appl. Phys. Lett. 104, 132910 (2014).

18. L. Fang, L. You, Y. Zhou, P. Ren, Z. S. Lim, and J. Wang, *Appl. Phys. Lett.* **104**, 142903 (2014).

19. R. K. Katiyar, P. Misra, F. Mendoza, G. Morell, and R. S. Katiyar, Appl. Phys. Lett. 105, 142902 (2014).

20. R. K. Katiyar, Y. Sharma, P. Misra, V. S. Puli, S. Sahoo, A. Kumar, J. F. Scott, G. Morell, B. R. Weiner, and R. S. Katiyar, Appl. Phys. Lett. 105, 172904 (2014).

21. R. L. Gao, H. W. Yang, Y. S. Chen, J. R. Sun, Y. G. Zhao and B. G. Shen, Appl. Phys. Lett. 104, 031906 (2014).

22. R. L. Gao, H. W. Yang, Y. S. Chen, J. R. Sun, Y. G. Zhao and B. G. Shen, E. Phys. Lett. 105, 37008 (2014).

23. P. Maksymovych, S. Jesse, P. Yu, R. Ramesh, A. Baddorf, and S. Kalinin, *Science* **324**, 1421 (2009).

24. R. K. Katiyar, Y. Sharma, D. Barrionuevo, S. Kooriyattil, S. P. Pavunny, J. S. Young, G. Morell, B. R. Weiner, R. S. Katiyar, and J. F. Scott, *Appl. Phys. Lett.* **106**, 082903 (2015).

25. M. M. Shirolkar, C. Hao, X. Dong, T. Guo, L. Zhang, M. Li, and H. Wang, *Nanoscale* **6**, 4735 (2014).

26. J. Wu and J. Wang, *J. Appl. Phys.* **107**, 034103 (2010),

27. S. Hussain, S. K. Hasanain, G. H. Jaffari, and S. I. Shah, *C. Appl. Phys.* **15**, 194 (2015).

28. A. Gruverman, B. J. Rodriguez, A.I. Kingon, R. J. Nemanich, A. K. Tagantsev, J. S. Cross, and M. Tsukada, *Appl. Phys. Lett.* **83**, 728 (2003).

29. Y. Guo, B. Guo, W. Dong, H. Li, and H. Liu, *Nanotechnology* **24**, 275201 (2013).

30. S. Hong, J. Woo, H. Shin, J. U. Jeon, Y. E. Pak, E. L. Colla, N. Setter, E. Kim, and K. No, J. Appl. Phys. 89, 1377 (2001).

31. T.-H. Yang, Y.-W. Harn, K.-C. Chiu, C.-L. Fan, and J.-M. Wu, *J. Mater. Chem.* **22**, 17071 (2012).

32. C. Wang, K.-J. Jin, Z.-T. Xu, L. Wang, C. Ge, H.-B. Lu, H.-Z. Guo, M. He, and G.-Z. Yang, *Appl. Phys. Lett.* **98**, 192901 (2011).